# DESIGN AND ASIC IMPLEMENTATION OF DUC/DDC FOR COMMUNICATION SYSTEMS

Naagesh S. Bhat[1]

[1]Senior Product Engineer, Green Mil International Ltd., Bangalore, India
bnsnagesh@gmail.com

## ABSTRACT

*Communication systems use the concept of transmitting information using the electrical distribution network as a communication channel. To enable the transmission data signal modulated on a carrier signal is superimposed on the electrical wires. Typical power lines are designed to handle 50/60 Hz of AC power signal; however they can carry the signals up to 500 KHz frequency. This work aims to aid transmission/reception of an audio signal in the spectrum from 300 Hz to 4000 Hz using PLCC on a tunable carrier frequency in the spectrum from 200 KHz to 500 KHz. For digital amplitude modulation the sampling rate of the carrier and the audio signal has to be matched. Tunable carrier generation can be achieved with Direct Digital Synthesizers at a desired sampling rate. DSP Sample rate conversion techniques are very useful to make the sampling circuits to work on their own sampling rates which are fine for the data/modulated-carrier signal's bandwidth. This also simplifies the complexity of the sampling circuits. Digital Up Conversion (**DUC**) and Digital Down Conversion (**DDC**) are DSP sample rate conversion techniques which refer to increasing and decreasing the sampling rate of a signal respectively.*

*The objective was to design and implement low power ASIC of DUC and DDC designs at 65nm for PLCC. Low power implementation was carried out using Multi-VDD technique. MATLAB software models were used to understand the DUC and DDC designs. RTL to GDS flow was executed using Synopsys tools such as VCS, Design Compiler, IC Compiler and PrimeTime. Key milestones of this activity are RTL verification, synthesis, gate-level simulations, low power architecture definitions, physical implementation, ASIC signoff checks and postroute delay based simulations. Multi-VDD technique deployed on DUC and DDC helped to reduce the power consumption from 280.9uW to 198.07uW and from 176.26uW to 124.47uW respectively. DUC and DUC designs have met functionality at 64MHz clock frequency. Both the designs have passed postroute delay based simulations, static performance checks, power domain checks and TSMC's 65nm design rule checks.*

## 1. INTRODUCTION

This paper focuses on Design and ASIC Implementation of Digital up-converter and Down converter for communication applications at 65nm technology. Digital up-conversion and down-conversion are well known sample rate conversion processes in Digital Signal Processing. These techniques are widely used for converting a baseband signal to band pass signal and vice-versa to enable the transmission and reception. For the baseband signal to be transmitted, it needs to be modulated on to an IF/RF carrier frequency. Nyquist theorem [1] says the sampling rate shall be at least twice the highest frequency component. Hence the base band signal, whose sample rate might be very less compared to IF/RF carrier signal sampling rate, needs to have the sampling rate to match the IF/RF carrier signal sampling rate. It's the reverse process with respect to receivers. In case of receivers the sample reduction helps to reduce the processing complexity of the received baseband signal. In simple, down conversion can be defined as removing samples (also called as Decimation) and generating new samples by virtue of adding zeroes (also called as Interpolation) and interpolate the new samples. Basic sample rate conversion is explained in the Figure 1 and Figure 2.

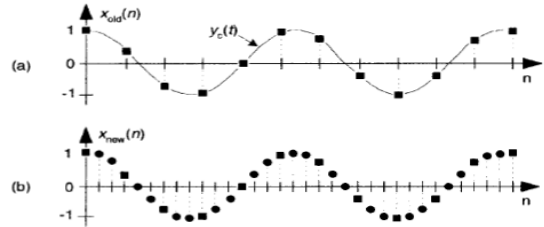

Figure 1. Interpolation example: (a) original sequence (b) interpolate by four sequence

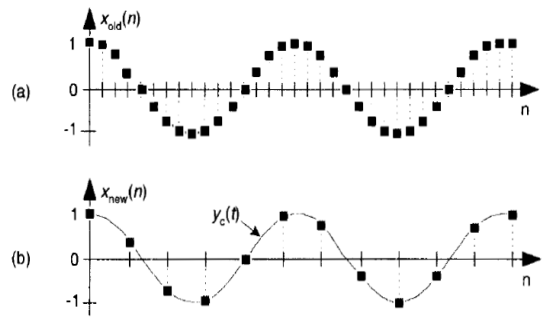

Figure 2. Decimation example: (a) original sequence (b) interpolate by four sequence

In a PLCC system the communication is established through the power line. The DUC aims to aid the transmission of the audio signal in the spectrum between 300Hz to 4000Hz, on a carrier frequency which can vary from 200 KHz to 500 KHz. The DUC system gets its input from a 14-bit ADC at a sample rate of 64 KHz. The carrier signal generated by a DDS (Direct Digital Synthesizer) is at a sample rate of 1.28MHz. This forms the specification of the DUC to up sample the audio signal by 20. DUC design output is fed to a Digital to Analog Converter (DAC) whose output is then super imposed on a power line for transmission. Since DDC is considered as a part of receiver, its objectives are just opposite to DUC.

CIC filters are good choice for implementing decimation or interpolation because they don't use multipliers and their frequency response can reduce aliasing and imaging issues resulting due to decimation and interpolation respectively [3]. However non-flat pass band response of CIC filters need a compensation filter either before or after the CIC filter [4] to compensate the gain loss. Unlike conventional signal generation, Direct Digital Synthesizers enable micro-Hertz tuning resolution, extremely fast frequency hopping, digital control interface and elimination of manual tuning to tweak the performance [5]. DDS implementation is very simple; it can be built using a phase accumulation circuitry and a look-up table preserving the signal samples. Having DDS on chip avoids the need of sampling circuits and provides a great flexibility in tuning to the required frequency.

Section 2 describes the specifications of DUC and DDC designs. Section 3 shows the software models constructed for DUC and DDC. Section 4 describes the ASIC system design details of DUC and DDC. Section 5 shows the RTL verification carried out and the results. Section 6 discusses the low power implementation of DUC and DDC. Section 7 discusses ASIC implementation and the results achieved. Section 8 provides conclusions of this work.

## 2. DESIGN SPECIFICATIONS

Functional and technology specifications of DUC and DDC designs are described in this section.

### 2.1. DUC

The up conversion system gets its input from an ADC with 14-bit resolution, 0-5V range and a sampling frequency of 64 KHz. The input signal spectrum is from 300 to 4000Hz. The input signal has to be up-sampled by 20 and get mixed with a carrier signal ranging from 200 KHz to 500 KHz. The Direct Digital Synthesizer which generates the carrier signal is also expected to be part of the up-sampling system.

### 2.2. DDC

The down conversion system gets its input from an ADC with 14-bit resolution, 0-5V range and a sampling frequency of 1.28 MHz. The input signal spectrum is from 200 KHz to 500 KHz. Demodulation is expected to be part of the down conversion system. Hence the input signal is expected to be mixed with a carrier signal to generate the baseband signal. The baseband signal then has to be decimated by 20 to get the sample rate of 64 KHz.

### 2.3. Technology Specifications

| | | |
|---|---|---|
| Foundry | : | TSMC |
| Technology | : | 65nm |
| Process | : | 0.5 to 1.5 (0.5 = Best and 1.5 = Worst) |
| Operating Voltage | : | 1.1V (1.08V = Worst and 1.32V = Best) |
| Operating Temperature | : | -40C to 125C |
| Master Clock Frequency | : | 64 MHz |
| Internal generated clocks | : | 1280 KHz and 64 KHz. |
| IO timing | : | 50-50 basis |

## 3. DESIGN OF SOFTWARE REFERENCE MODELS

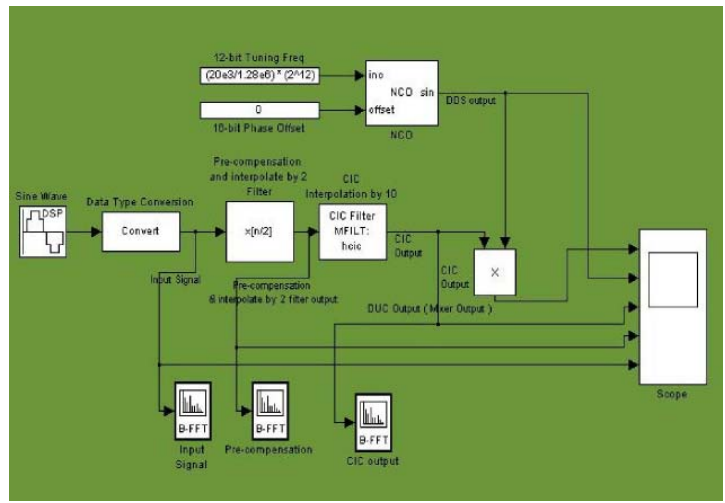

Figure 3. DUC Software model

DUC and DDC software reference models were constructed to understand the functionality. As this paper focuses on ASIC low power implementation, a detailed study on coefficient calculation, and magnitude response of the filters or arithmetic considerations were not considered. Focus is to design a code which can be configured as per system clock and filter coefficients. Software models were extremely useful in RTL verification as well.

Figure 3 and Figure 4 show the software models of DUC and DDC systems respectively. Converter is used to pick up 14-bit input. The compensation filters are simple MAC (multiply and accumulate) based filters which act to compensate for the non flat pass band of CIC filter used for interpolation/decimation, which also does interpolation/decimation by two. Both the CIC filter and the compensation filter were designed using FDATool. The Direct-Digital Synthesizer is constructed using an NCO (Numerically Controlled Oscillator). The output of Sine wave from DDS is considered with 1V magnitude by default.

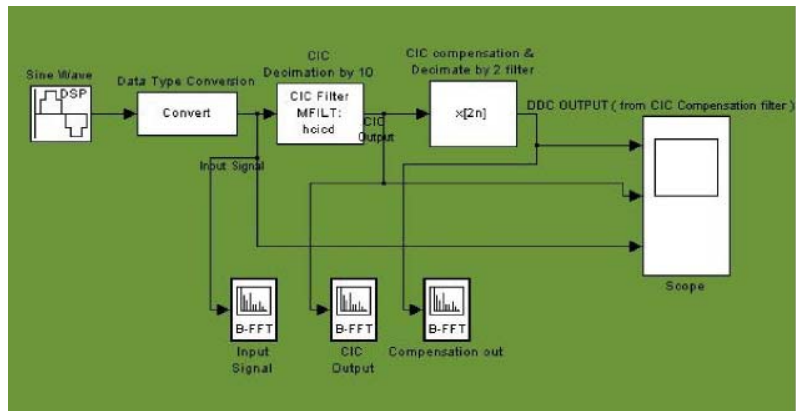

Figure 4. DDC Software Model

Figure 5 and Figure 6 show the simulation results of the software models at each stage of the DUC and DDC systems respectively.

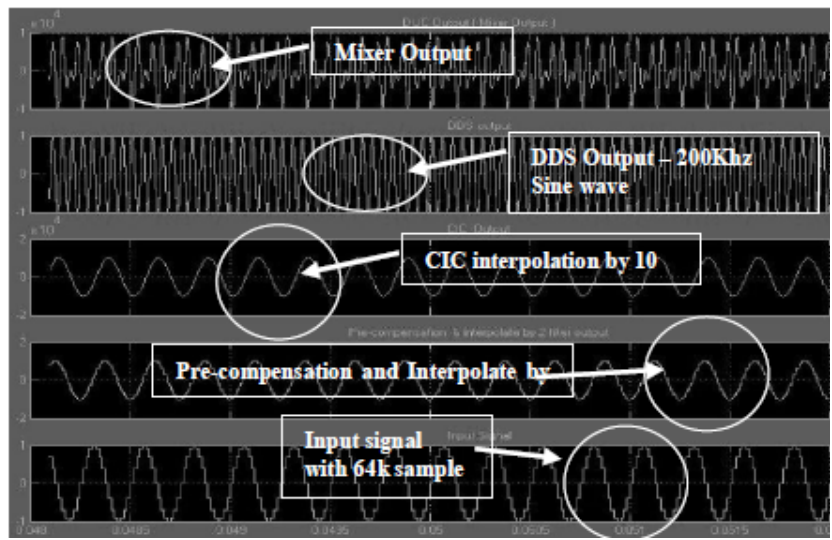

Figure 5. DDC System Analysis using Software Model

DUC system is simulated with an input Sine wave of 4 KHz frequency with a sample rate of 64 KHz, 16 samples per Sine wave. Compensation cum interpolate-by-2 filter increases the sample rate to 128 KHz, 32 samples per Sine wave. The CIC interpolation filter increases the sample rate by 10 to reach 1280 KHz sample rate, now the output of CIC filter contains 320 samples per Sine wave. This demonstrates the successful sample rate conversion from 64 KHz to 1280 KHz, the specification of DUC system. DUC system also supports mixing of the up-sampled input signal with a carrier frequency. The DDS/NCO is programmable such that it can generate frequencies between 200 KHz to 500 KHz. In the software model DDS is setup to generate 200 KHz Sine wave, with a sample rate of 1280 KHz. It is a must to have the sample rate of the output of DDS equal to the sample of rate of signal with which it is mixed. As shown in Figure 5 Sine wave is successfully generated by DDS, which gets mixed with the output of CIC filter, which is already at 1280 KHz. Figure 5 gives a pictorial explanation of the change in sample rate, change in number of sample samples per Sine wave, DDS output and mixer operation.

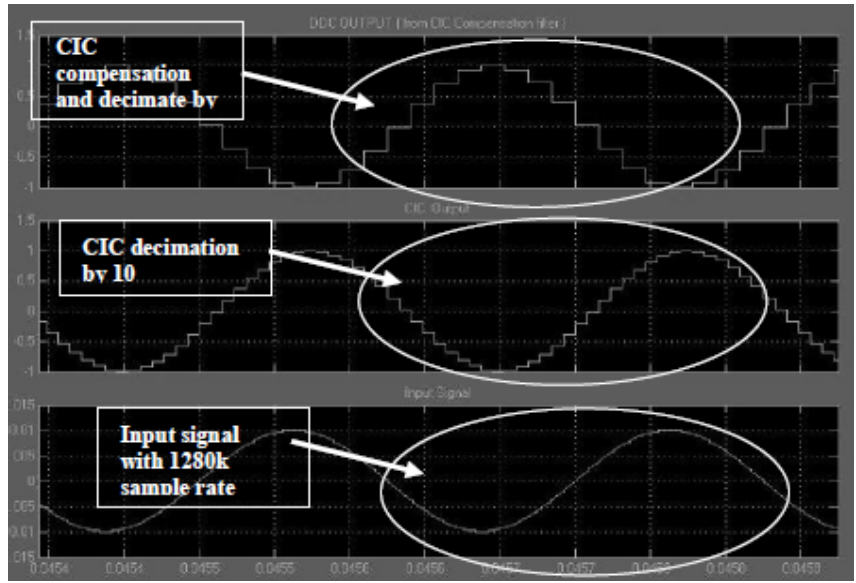

Figure 6. DDC System Analysis using Software Model

For simplification purpose the mixer part of the DDC is not considered in software model. The DDC system is simulated with a 4 KHz Sine wave assumed to be the output of the mixer at a sample rate of 1280 KHz, 320 samples per Sine wave. The CIC decimation filter decreases the sample rate by 10 to reach 128 KHz sample rate, now the output of CIC filter contains 32 samples per Sine wave. Compensation cum decimate-by-2 filter decreases the sample rate to 64 KHz, 16 steps per Sine wave. This demonstrates the successful sample rate conversion from 1280 KHz to 64 KHz, the specification of DDC system. Figure 6 gives a pictorial explanation of the change in sample rate and change in number of sample samples per Sine wave.

## 4. ASIC SYSTEM DESIGN

### 4.1. DUC

DUC ASIC architecture as in Figure 7 is similar to that of shown MATLAB software mode in Figure 3; however an additional mixer is added prior to the interpolation. A constant 20 KHz carrier is considered to mix with the data so that the filters which follow in the system need not have a steep transition band and hence less number of coefficients. A high pass filter immediately follows the constant frequency mixer to select the upper band (20 KHz + signal frequency) output from the mixer. Another change in the system is that compensation filter doesn't implement any interpolation; the CIC interpolation filter itself does the interpolation by 20 functions. Clock generation is an obvious addition in ASIC system.

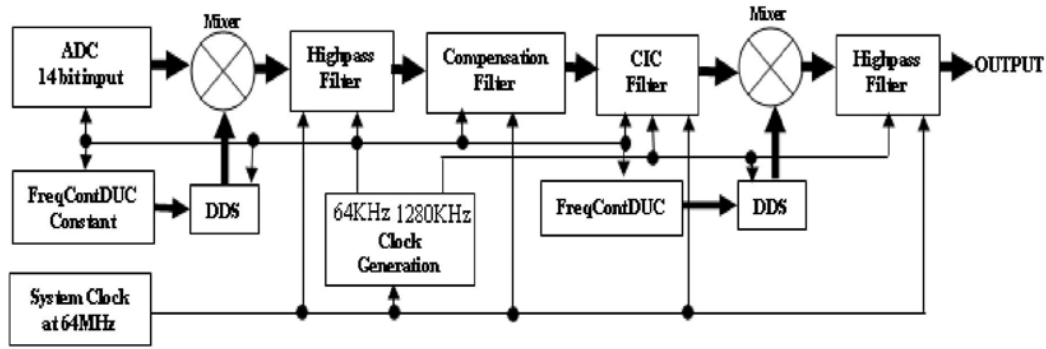

Figure 7. DUC ASIC Architecture

## 4.2. DDC

DDC ASIC architecture as in Figure 8 is similar to that of shown MATLAB software mode in Figure 4; however the demodulation mixer is added prior to the decimation. The DDS is programmable and generates the carrier frequency between 200 KHz to 500 KHz. A high pass filter immediately follows the constant frequency mixer to select the upper band (carrier + signal frequency) output from the mixer. Another change in the system is that compensation filter doesn't implement any decimation; the CIC decimation filter itself implements decimate by 20 function.

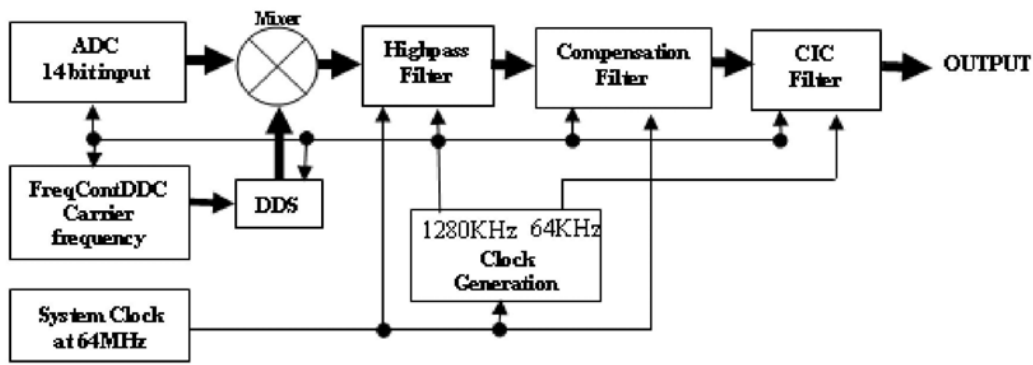

Figure 8. DDC ASIC Architecture

## 4.3. RTL Improvement - Clock

When these designs were implemented on FPGA, DCM (Digital Clock Manager) was deployed to generate 64 KHz and 1280 KHz clocks. Source clock was 100 MHz. For ASIC implementation 64 MHz clock is chosen as master clock, from which 64 KHz and 1280 KHz clocks are generated by dividing the master clock by 1000 and 50 times respectively. The 100MHz clock in FPGA design also works as internal clock in executing the MAC operations (Multiply and accumulate). Section 4.4 explains the bug in the design in MAC operation leading the reason for choosing 64 MHz as master clock. These benefit in power savings also.

### 4.4. MAC Bug and Fix

MAC is implemented such that the input stream comes at lower frequency and the multiplication and accumulation operation happens at a higher frequency to make the system data flow stream continuous. Input to the MAC is stored in a shift register setup, where the shift register is read into MAC operation by a slower clock and the MAC operation is implemented by a faster clock. However the slower and faster clocks need to be synchronized to make the MAC operation effective. Having an integer relationship helps a lot in static timing analysis, by avoiding LCM related concerns in case of non-integer clocks.

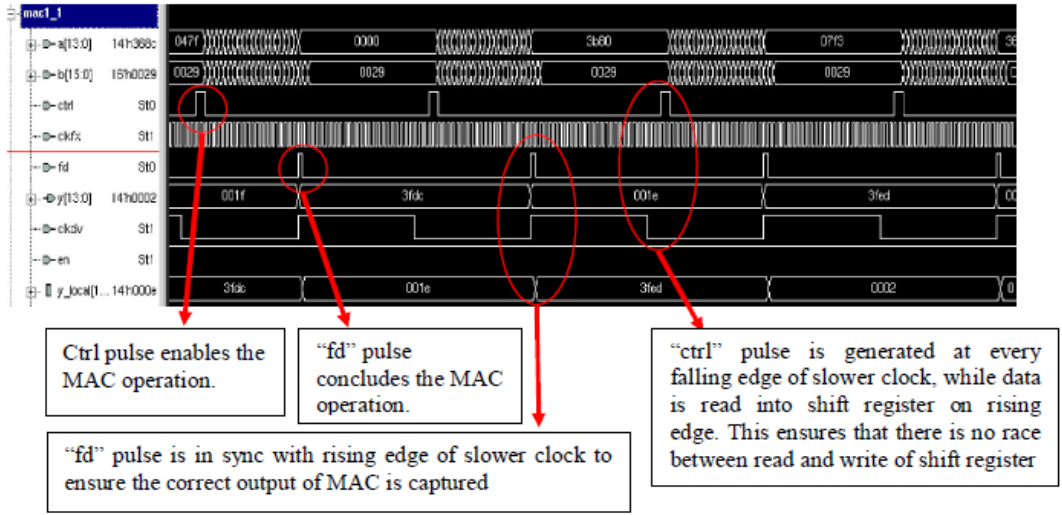

Figure 9. MAC Operation Synchronization

The synchronization issue has been corrected with control signals generation ("**ctrl**" and "**fd**"). The control signals are generated from the clock division circuits such that the faster clock collects the data only after slower clock feeds the data into shift register and the vice-versa to collect the output from MAC operation. Figure 9 shows the control signal generation and the how they control the MAC operation.

For ASIC implementation 24 coefficients are used which make the 64 MHz a good choice. 24 multiplications with accumulation need 25 clock cycles. Since 1280 KHz clock frequency is higher compared to 64 KHz, considering half-cycle of 1.28 MHz clock available for MAC operation, the system clock can be derived as 64 MHz (1280*10^3*(24+1)*2)

### 4.5. Other Code Changes

For coefficient storage, initial statements were used in the code for FPGA implementation. However, initial statements are not governed by synthesis tools. Hence the coefficient storage and lookup tables are implemented using case constructs.

Added synchronous reset controls to the DUC and DDC systems.

As explained in the Section 4.4, corrected the MAC code to make it functional. MAC correction also led to system frequency configurability. Accordingly master clock frequency is reduced and the necessary internal clock generation logic implemented. FPGA design used DCMs for clock generation.

Enabled the code to register the input/output data and the frequency settings, made this in synchronous with 64 KHz clock in case of DUC system and 1280 KHz clock in case of DDC system. It's the vice-versa for the output data.

System operation is made to be on hold, till the clock generation is completed. The clock generation circuit emits a control signal to start the DUC/DDC operation.

Initialization of various signals/registers in the design were implemented to enable the gate level synthesized netlist to pass the simulations, this was not a problem for FPGA design. This also enabled to have the logic un-operational till the clocks are generated and also the logic to react to reset of the system.

## 5. RTL VERIFICATION

### 5.1. Clock Generation

64 MHz clock is considered as source to generate 64 KHz and 1280 KHz. To generate 64 KHz clock, a division of 1000 is applied. Similarly to generate 1280 KHz clock a division of 50 is applied on 64 MHz clock. Figure 10 shows the output of VCS simulation, it shows/proves the relationship between 64 KHz clock and 1280 KHz clock, a count of 20 pulses of 1280 KHz clock can be observed under 64 KHz clock waveform.

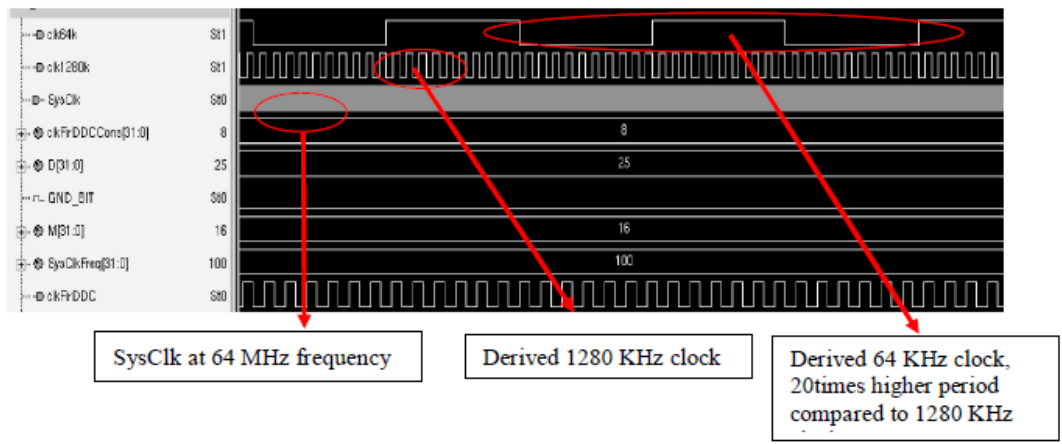

Figure 10. Clock Generation for DUC and DDC

### 5.2. DDS Functioning

DDS implementation is carried out using a lookup table which is loaded with two's complement values representing the Sine wave. The frequency number that is set as input determines the number of elements skipped consequently in the lookup table. Figure 11 shows the VCS simulation output for a frequency of 105 KHz. Figure 11 shows the phase accumulation which is in digital, and the sine wave output in analog. This is one of the critical parameters to be met in implementing Direct Digital Synthesizer. Size of the lookup table is 256 with eight bit words. In the VCS simulation result shown in Figure 11, the reference clock used is of 1280 KHz.

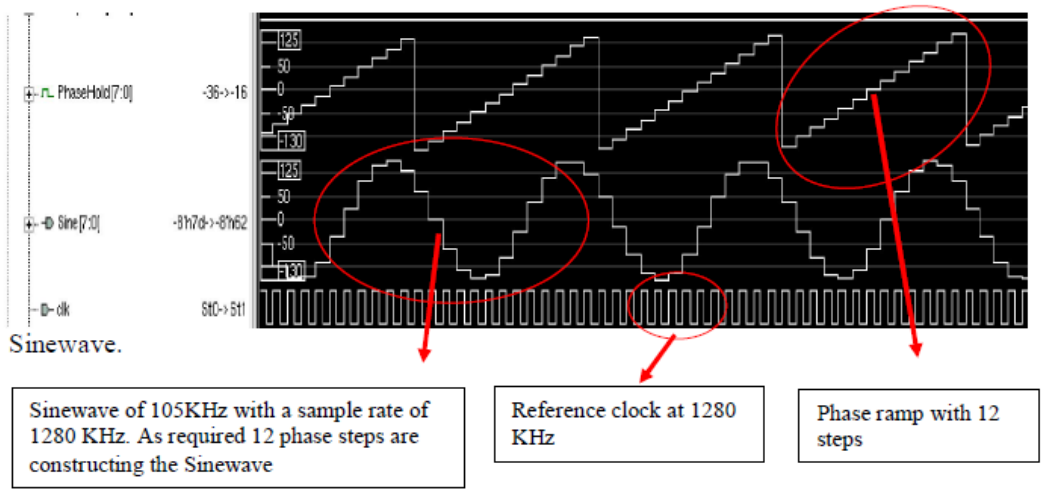

Figure 11. DDS Functioning

### 5.3. Mixer

Digital mixer is used in both DUC and DDC systems. In DUC it is used to modulate input signal with 20 KHz intermediate carrier and also to convert the final up sampled data on to the IF carrier (varies from 200 KHz to 500 MHz). Figure 12 shows the waveforms representing the mixer operation. Input data stream is represented by "a" (of 14 bit) and the digital sine wave is represented by "b" (of 8 bit). Output of the mixer operation is represented by "P". Multiplication operation is completed in one clock cycle as can be seen in the Figure 12.

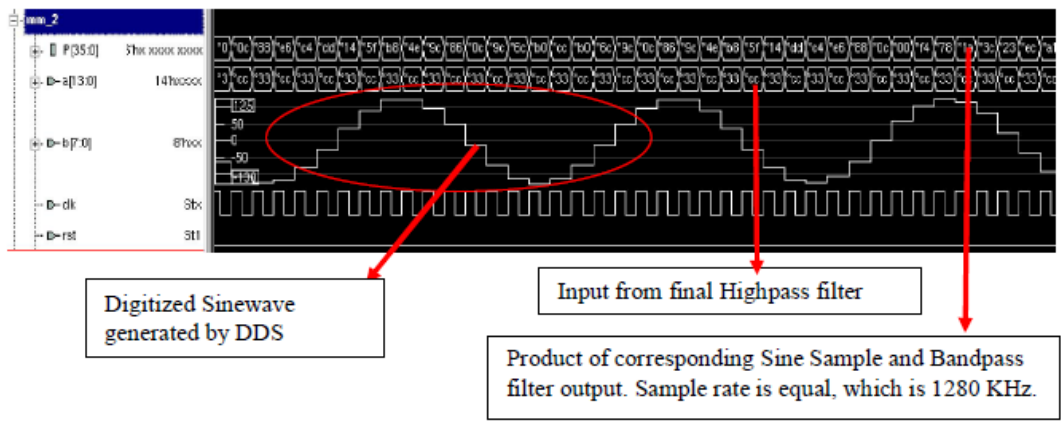

Figure 12. Mixer Operation

### 5.4. MAC Operation

Multiply and accumulate (MAC) operation is very crucial in the filter operation. The MAC is implemented such that the input stream comes at a frequency of either 64 KHz or 1280 KHz (for DDC and DUC respectively) and the MAC operation actually runs at 64 MHz. This is carried out intentionally to continue the stream of data flow uninterrupted and also to avoid any need of additional memory and processing in holding the data for MAC operation.

The MAC operations are controlled by the "**ctrl**" and "**fd**" signals as shown in Figure 13. These signals make the MAC operation synchronous to either 64 KHz or 1280 KHz frequency depending on the DDC or DUC application respectively. "**ctrl**" signal initializes the MAC operation and "**fd**" signal concludes the MAC operation taking the final output to the register "**y**".

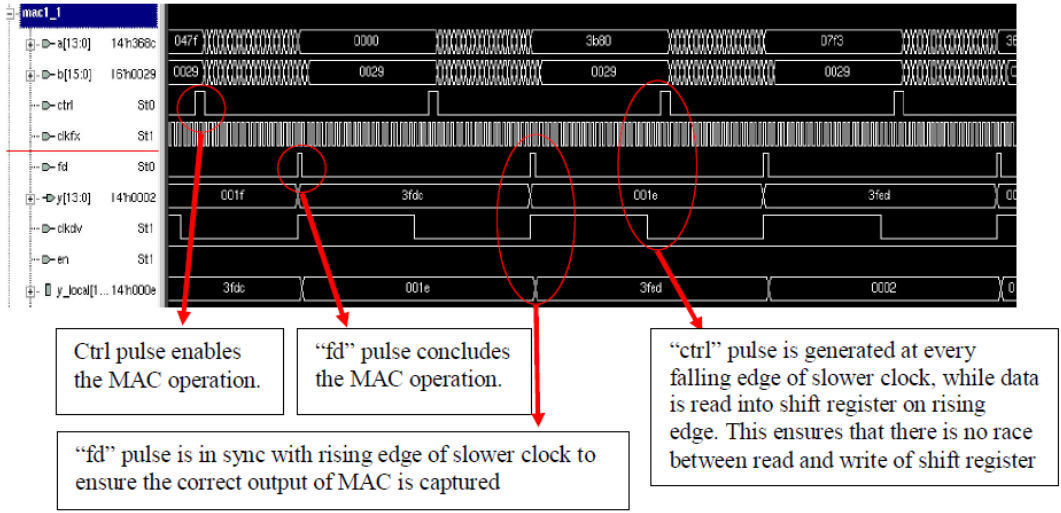

Figure 13. MAC Operation

## 5.5. Highpass Filter

Highpass filters implemented are simple MAC based ones. Figure 14 shows the input data stream from mixer and coefficients mapped. The input "**x**" is actually stored in a shift register and passed on to MAC. "**ctrl**" and "**fd**" signals act as control signals to initiate the MAC operation and capture the MAC output respectively. As described in Section 5.4, the MAC operation works at higher frequency where as input/output are sent/captured by slower clock "**fd**" control signal synchronizes and helps to capture the output of the filter operation.

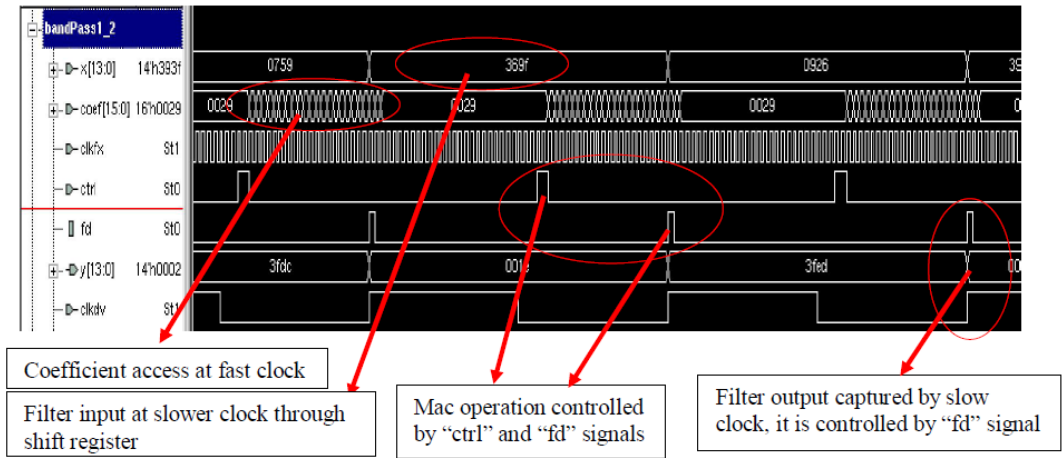

Figure 14. Filter Operation with MAC

## 5.6. CIC Decimation

The CIC filter in DDC system implements decimate by 20 to convert the sample rate of the input signal from 64 KHz to 1280 KHz. The integrator section of the filter works at 1280 KHz and the differentiator section of the filter works at 64 KHz. In Figure 15 signal_in_ext is the input data coming at 1280 KHz frequency. Integrator and differentiator shown in Figure 15 reflect the result of integration and differentiation actions performed. Integrator works at the sample rate of 1280 KHz. Output "y" from differentiator is produced at a sampling rate of 64 KHz concluding the decimation operation.

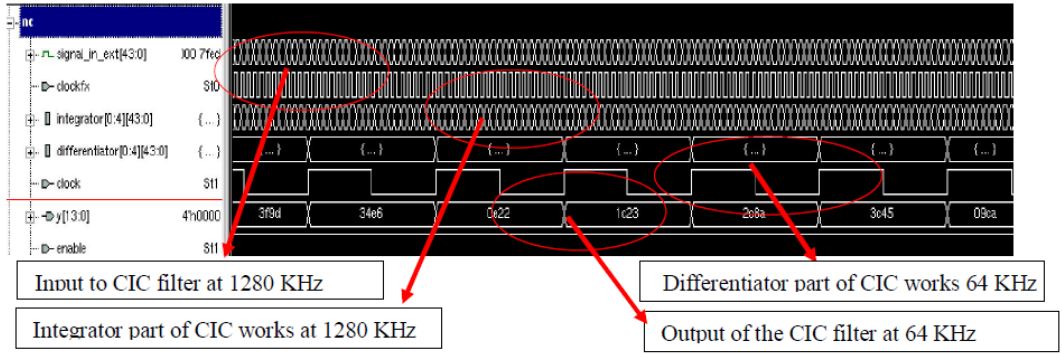

Figure 15. CIC Decimation

### 5.7. CIC Interpolation

The CIC filter in DUC system implements the interpolation by 20 to convert the sample rate of the input signal from 64 KHz to 1280 KHz. The integrator section of the filter works at 1280 KHz and the differentiator section of the filter works at 64 KHz. In Figure 16 signal_in_ext is the input data at 64 KHz frequency. Integrator and differentiator shown in Figure 16 reflect the result of integration and differentiation actions performed. Output "y" from integrator is produced at a frequency of 1280 KHz concluding the interpolation operation.

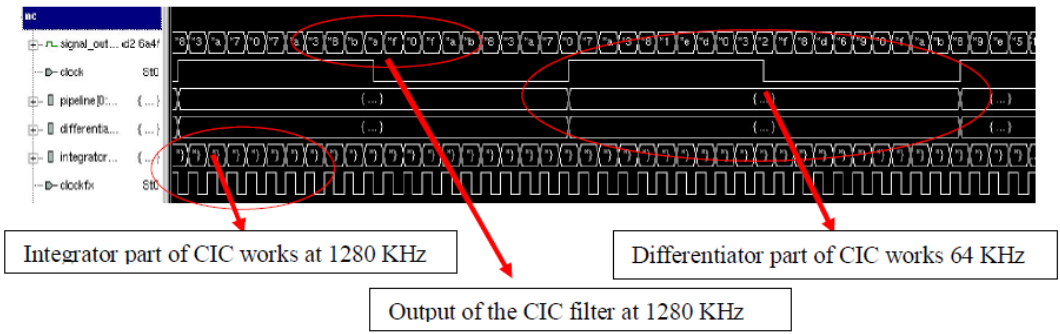

Figure 16. CIC Interpolation

### 5.8. DDC Operation

The sections till 5.7 explain how the subsystems of the DDC work. The integrated system output is shown in Figure 17. The test bench supplies input at 1280 KHz sampling rate and the output is observed at 64 KHz, proving the basic functionality of the DDC system per specification.

System operation is initiated with a reset and the system waits till the 64 KHz and 1280 KHz clocks are generated. The clock divider circuit initiates an enable signal to start capturing the input data and there by initiates the digital down conversion process. The clock divider takes 500 clock cycles of 64 MHz master clock to generate 64 KHz clock, till then the system doesn't capture any input. The below are the other latencies in the system,

1. Data registering takes one clock cycle of 1280 KHz clock.

2. Mixer operation takes one clock cycle of 1280 KHz clock

3. Highpass filter after mixer takes one clock cycle of 1280 KHz clock.

4. Compensation filter takes one clock cycle of 1280 KHz clock.

5. 5 stage CIC filter integrator section takes five clock cycles of 1280 KHz clock.

6. 5 stage CIC filter differentiator section takes five clock cycles of 64 KHz clock.

All together the DDC system takes (500 clock cycles of 64 MHz clock + 9 clock cycles of 1280 KHz clock + 5 clock cycles of 64 KHz clock) to produce the output. Figure 18 gives a pictorial explanation of the same.

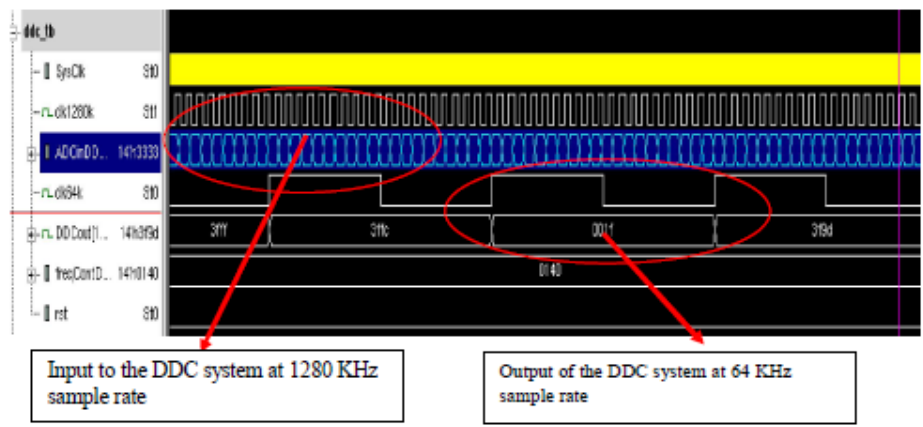

Figure 17. DDC System Operation

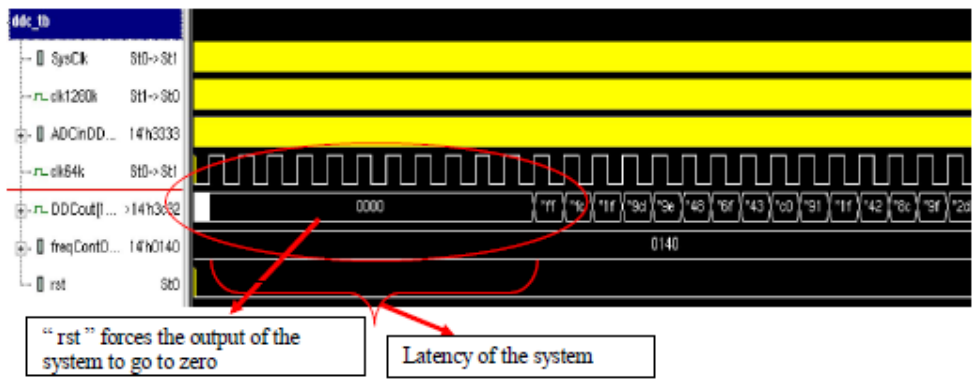

Figure 18. DDC System Latency

## 5.9. DUC Operation

The integrated DUC system output is shown in Figure 19. The test bench supplies input at 64 KHz sample rate and the output is observed at 1280 KHz, proving the basic functionality of the DUC system per specification. System operation is initiated with a reset and it waits till the 64 KHz and 1280 KHz clocks are generated. The clock divider circuit initiates an enable signal to start capturing the input data and there by initiates the digital up conversion process. The clock divider takes 500 clock cycles of 64 MHz master clock to generate 64 KHz clock, till then the system doesn't capture any input. The below are the other latencies in the system,

1. Data registering takes one clock cycle of 64 KHz clock.

2. Mixer operation takes one clock cycle of 64 KHz clock

3. Highpass filter after mixer takes one clock cycle of 64 KHz clock.

4. Compensation filter after mixer takes one clock cycle of 64 KHz clock.

5. 5 stage CIC filter differentiator section takes five clock cycles of 64 KHz clock.

6. 5 stage CIC filter integrator section takes five clock cycles of 1280 KHz clock.

7. Mixing of carrier frequency takes one clock cycle of 1280 KHz clock.

8. Highpass filter after the above mixer takes one clock cycle of 1280 KHz.

All together the DUC system takes (500 clock cycles of 64 MHz clock + 9 clock cycles of 64 KHz clock + 7 clock cycles of 1280 KHz clock) to produce the output. Figure 20 gives a pictorial explanation of the same.

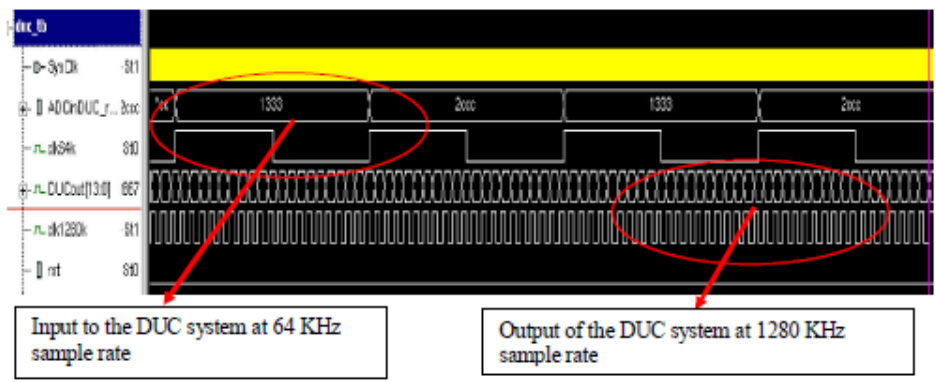

Figure 19. DUC System Operation

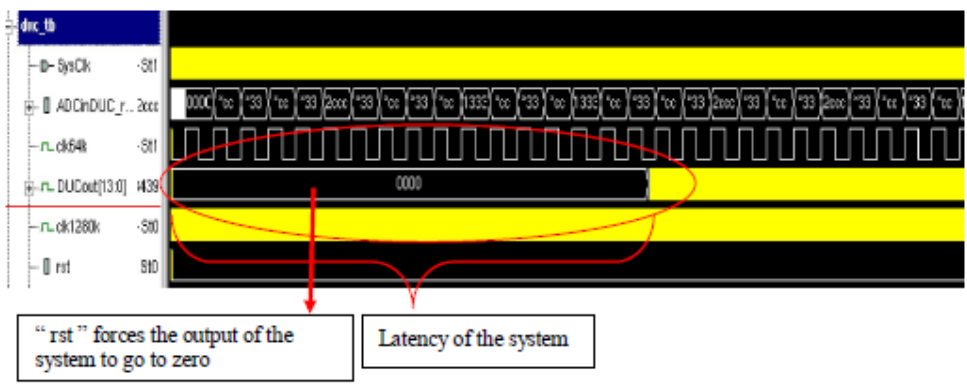

Figure 20. DUC System Latency

## 6. LOW POWER IMPLEMENTATION

### 6.1. Implementing Multi-VDD

Multi-VDD technique enables the entire DUC and DDC systems to work at 0.9V operating voltage, against the specification voltage of 1.08V. However the interface to these designs is still assumed to be at 1.08V. Figure 21 explains the power intent of DUC/DDC system pictorially. Level shifters are required to transfer data between 0.9V domain to 1.08V domain and vice-versa. Definitions of the domains, creating rules for level shifters, definition of power/ground nets etc are carried out.

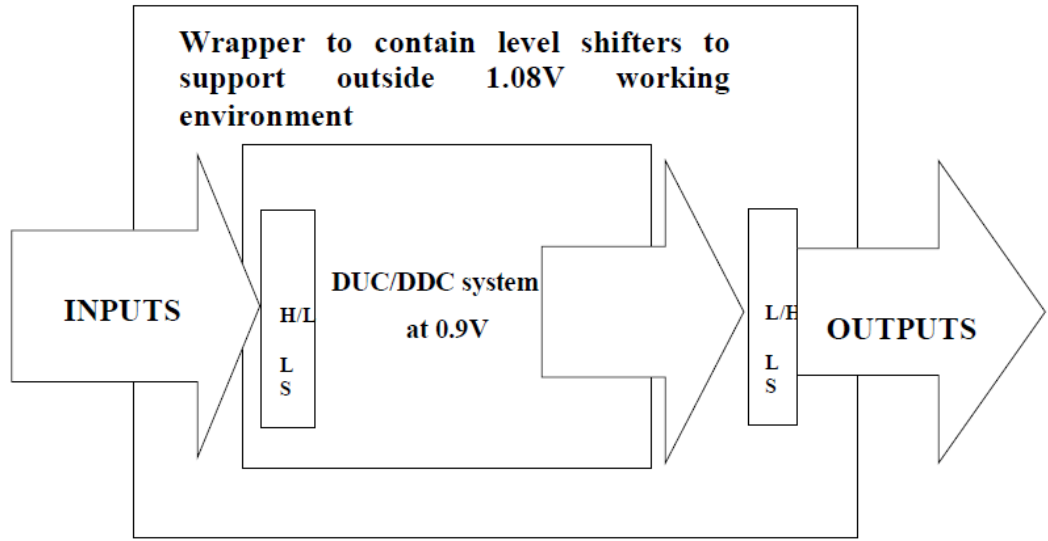

Figure 21. Low Power Architecture for DUC and DDC

### 6.2. Constraints Development

The DUC and DDC systems work on three clocks. 64 MHz, 64 KHz and 1280 KHz are the frequencies. 64 MHz is the input clock which acts like a master clock to generate 64 KHz and 1280 KHz clocks. By the virtue of data flow, all these three clocks are synchronous to each other. Even balancing the clock tree is carried out with respect to master clock, 64 MHz. Apart from these functional clocks, the design expects a scanclk for scanmode operation. False paths have been defined from scanclk to all functional clocks, although scanmode is also present to select the mode of design. Input ports of DUC design are constrained using 64 KHz clock and output ports of DUC design are constrained using 1280 KHz clock. 50% of clock period is assumed as input/output delay on the ports. Input ports of DDC design are constrained using 1280 KHz clock and output ports of DUC design are constrained using 64 KHz clock. 50% of clock period is assumed as input/output delay on the ports.

## 7. ASIC IMPLEMENTATION

### 7.1. Die Size Calculation

In 65nm designs the typical maximum utilization before going to routing can be up to 85-90%. For this design a budget of 20% is given for optimization, 3% for clock tree synthesis, and another 2% for hold fixes. Hence 60% is chosen as start utilization for place and route. Die size calculation for DDC and DUC designs are shown in Table 1.

Table 1. Die Calculation of DUC and DDC Design

| Die Size Calculation | DUC | DDC |
|---|---|---|
| Area from Synthesis report | 28819.079 um^2 | 37312.919 um^2 |
| P and R Start Utilization | 70% | 70% |
| Die area Required | 28819.079918 * 1.3 | 37312.91985 * 1.3 |
|  | 37464.80um^2 | 48506.8um^2 |
| Width / Length of the Core | 193.5um | 220.24um |
| Distance between IO to Core | 10um | 10um |
| Effective Die Height / Width | 193.5 + ( 10 * 2 ) | 220.24 + ( 10 * 2 ) |
|  | 213.5um | 240.24um |
| Rounding the Die Width / Height | **214um** | **241um** |

### 7.2. Floor Plan

Die size calculated in the Section 7.1 drives the floor plan definition using IC Compiler. Additional challenge in creating the floor plan is to define the power domains and able to control placement of cells, level-shifters and ensuring sanity during power domain crossings. Synthesis instantiates the level shifters up front, and passes on the power connection related information through Power Report to IC Compiler. IC Compiler understands power report and allows creating "voltage areas" and the "associated power and ground nets". Figure 22 shows the voltage areas, placement of level shifters and pins of the DUC/DDC designs. Both power domains use common ground; hence no ESD issues can be expected. The 0.9V voltage area is named as PD_1 domain and the 1.08V voltage area is named as PD voltage domain. The low to high level shifters sitting in PD domain need 0.9V supply also, they are routed to the 0.9V supply rails in PD_1 domain.

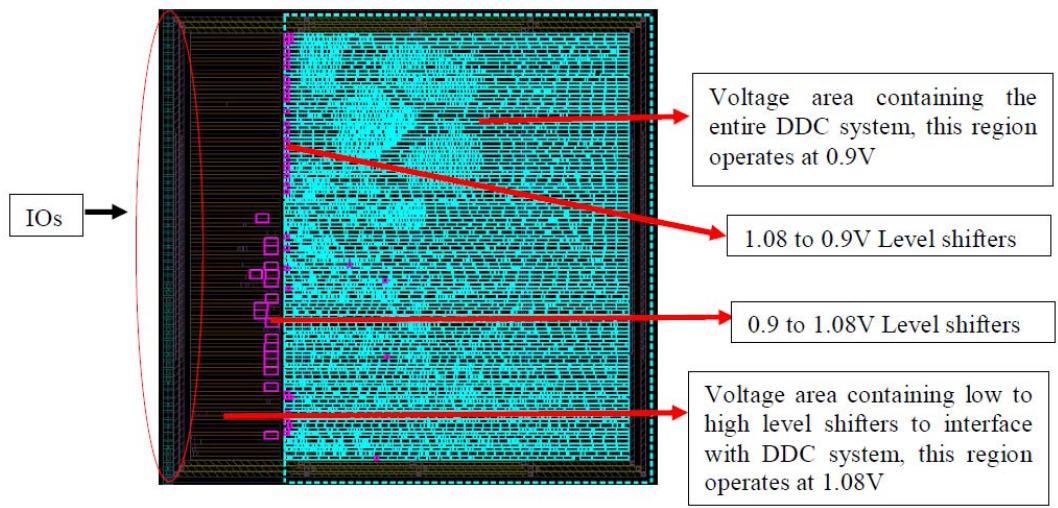

Figure 22. Voltage Islands in Floor Plan

### 7.3. Clock Tree Synthesis

Figure 23 shows the snapshot of Clock tree synthesis in DDC and DDC designs. Clock tree synthesis is very essential for the reasons

1. It is imperative to resolve the huge fan-out on the clock pin of the design. Else the clock network needs huge metal widths and huge driver to supply the current needed to drive all clock pins of the flops. The later options are not possible to implement.

2. While it is important to fix the above issue, it is equally important to synchronize the clock edges by reducing the delay between them. Resolving first issue means, inserting a buffer tree to balance the loads.

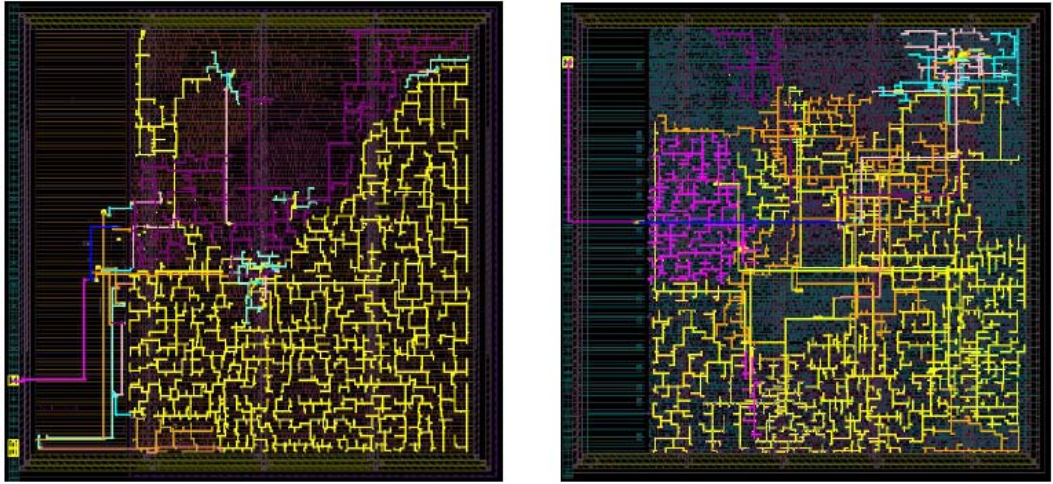

Figure 23. Clock Trees in DUC and DDC Designs

### 7.4. Routing

Routing script of IC Compiler is created such that it does timing-aware routing, fixes DRCs, leaves no LVS issues, does post route setup and hold timing optimization. Routing didn't experience any congestion, though there are many lookup tables in the design. Typically, the lookup table decoding logic creates huge connectivity. This makes the placer to create low density placement regions, which still wouldn't be routable due to heavy connectivity. Routing is observed successful with zero routing DRC and LVS violations. Figure 24 shows the snapshot of routing in DDC and DDC designs.

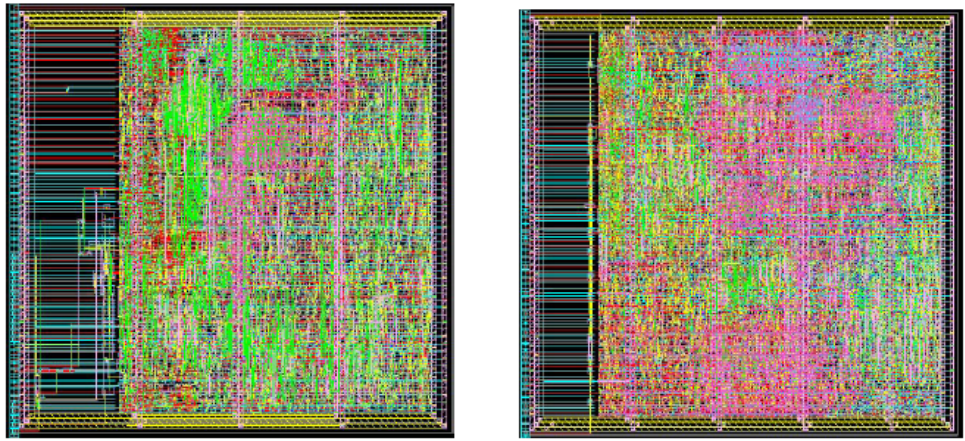

Figure 24. Routing Snaps of DUC and DDC Designs

### 7.5. Post-Route Simulations

The simulations run on RTL and Gate level netlist (after synthesis) work with unit delays. Post-route simulation is very important because the netlist goes through changes to meet performance requirements, clocks are no more ideal and on top of all these the setup and hold timing of the flops play an important role. It is obvious that post-route simulation needs to know the cell delays, net delays and the setup/hold time constraints. From STA tool we can write out a file in SDF format to represent these delays and constraints. SDF is known for Standard Delay Format, which represents cell delays, net delays and setup/hold timing requirements of the sequential cells. SDF files are generated from Prime Time tool using the post-route netlist and the parasitic extracted using IC Compiler. While reading the SDF and during the simulation, it is observed that VCS (simulation tool) hasn't complained any timing issues, assuring the timing performance of the design. The simulations results are observed to be satisfactory, proving that static timing analysis was complete for signoff. Figure 25 and Figure 26 show the post-route simulation results of DUC and DDC designs.

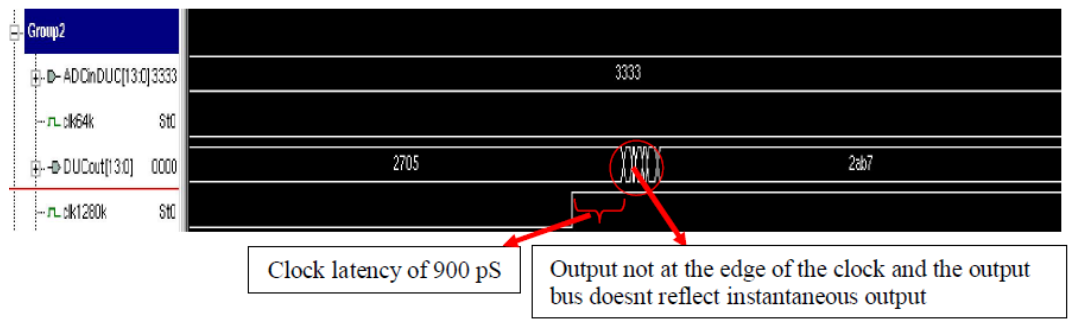

Figure 25. Post-route Simulation result of DUC

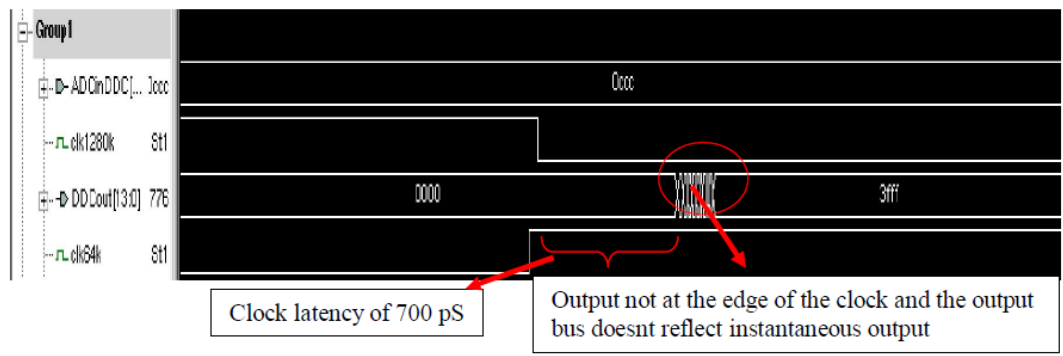

Figure 26. Post-route Simulation Result of DDC

### 7.6. ASIC Sign-off

ASIC signoff checks such as Static Timing Analysis (STA), physical verification and power domain checks were conducted. Timing has been met successfully in worst case and best corners. Functional simulations were found successful on post route netlist with cell and parasitic delays. This endorses the timing results achieved in STA. No DRC and LVS errors were found in the design.

## 8. CONCLUSION

Based on the work done on low power ASIC implementation of DUC and DDC and results achieved, the below listed conclusions are drawn:

- Multi-VDD low power technique is successfully deployed on DUC and DDC designs. In case of DDC the power consumption observed is 176.26uW with 1.08V as operating voltage, however with Multi-VDD (0.9V and 1.08V) technique the power consumption is reduced to 124.47uW. In case of DDC the power reduction is observed from 280.9uW to 198.07uW. Overall there is a 30% power reduction.

- RTL improvements are able to make the MAC operations to accept data at a slower clock and perform MAC operation at a faster clock frequency. These also helped to bring a relationship between the number of coefficients of a filter working at 1280 KHz frequency and the system clock frequency.

- RTL improvements also helped in reducing the power consumption by 35%

- Conventional clock gating technique deployed on DUC and DDC designs has helped to reduce about 11.5% and 8.5% power respectively

- Functional simulation results are found satisfactory on subsystems in the design like clock generation, mixer operation, DDS operation, MAC operation, MAC deployment in filters, CIC Interpolation and CIC decimation.

- Functional simulations on DDC and DUC show successful sample rate conversion by a factor of 20 as expected by the specification

- ASIC signoff criterion like Static Timing Analysis (STA), Physical Verification checks are successfully met